\documentclass[a4paper,fleqn,review]{cas-sc}

\usepackage[numbers]{natbib}
\usepackage{graphicx}%
\usepackage{url}
\usepackage{balance}
\usepackage{verbatim}
\usepackage{array}
\usepackage{stfloats}
\usepackage{textcomp}
\usepackage{tabularx}
\usepackage{makecell}
\usepackage{multirow}%
\usepackage{amsmath,amssymb,amsfonts}%
\usepackage{amsthm}%
\usepackage{mathrsfs}%
\usepackage[title]{appendix}%
\usepackage{xcolor}%
\usepackage{textcomp}%
\usepackage{manyfoot}%
\usepackage{booktabs}%
\usepackage{algorithm}%
\usepackage{algorithmicx}%
\usepackage{algpseudocode}%
\usepackage{listings}%
\usepackage{rotating}

\def\tsc#1{\csdef{#1}{\textsc{\lowercase{#1}}\xspace}}
\tsc{WGM}
\tsc{QE}
\tsc{EP}
\tsc{PMS}
\tsc{BEC}
\tsc{DE}

\begin{document}
\let\WriteBookmarks\relax
\def\floatpagepagefraction{1}
\def\textpagefraction{.001}

\shorttitle{ECHOv2: Frequency-Structured Pre-trained Acoustic Representation Model for Machine ASD}
\shortauthors{Zhang et~al.}

\title [mode = title]{ECHOv2: A Frequency-Structured Pre-trained Acoustic Representation Model with Cross-Band Modeling for Machine Anomalous Sound Detection}


\author[1,3]{Yucong Zhang}
\ead{yucong.zhang@whu.edu.cn}

\credit{Conceptualization, Methodology, Software, Investigation, Writing - Original Draft}

\author[2,1]{Juan Liu}
\ead{liujuan@whu.edu.cn}

\credit{Supervision, Project Administration, Writing - Review \& Editing}

\author[3,2]{Ming Li}
\cormark[1]
\ead{mingli369@cuhk.edu.cn}

\credit{Supervision, Project Administration, Writing - Review \& Editing}


\affiliation[1]{
    organization={School of Computer Science, Wuhan University},
    city={Wuhan},
    postcode={430072},
    state={Hubei},
    country={China}
}

\affiliation[2]{
    organization={School of Artificial Intelligence, Wuhan University},
    city={Wuhan},
    postcode={430072},
    state={Hubei},
    country={China}
}

\affiliation[3]{
    organization={School of Artificial Intelligence, The Chinese University of Hong Kong, Shenzhen},
    city={Shenzhen},
    postcode={518172},
    state={Guangdong},
    country={China}
}

\cortext[cor1]{Corresponding author}

\begin{abstract}
Machine anomalous sound detection (ASD) is an important technology for industrial acoustic monitoring, where robust acoustic representation learning remains challenging due to limited anomalous samples and complex machine sound characteristics.
Existing pre-trained acoustic representation models do not fully capture frequency-specific characteristics of machine sounds.
To address this, we propose ECHOv2, a frequency-structured pre-trained acoustic representation model. The model learns localized intra-band representations to capture fine-grained spectral patterns while also incorporating a two-level self-distillation strategy with explicit inter-band supervision to model cross-frequency dependencies.
The inter-band branch performs global context alignment and masked sub-band reconstruction, and multiple summary tokens are introduced for structured aggregation with controllable frequency granularity, enabling region-aware interaction across sub-bands during training.
This design allows ECHOv2 to robustly handle diverse machine types and noisy operating conditions while maintaining stable representation quality.
To enable fair and consistent evaluation of pre-trained audio backbones, we establish a unified ASD benchmark over DCASE 2020--2025 with two complementary protocols: embedding-based evaluation for frozen representation discriminability and adaptation-based evaluation for downstream transferability.
Ablation studies confirm the effectiveness of intra-band learning, inter-band supervision, and structured aggregation granularity for robust ASD representation learning.
These findings demonstrate that structured cross-band modeling improves acoustic representation learning for machine monitoring and provides an effective pre-trained representation model for industrial anomalous sound detection.
The model and benchmark are publicly available to promote reproducible research.
\end{abstract}




\maketitle


\section{Introduction}

Machine Anomalous Sound Detection (ASD) aims to identify abnormal machine conditions from acoustic observations and plays an important role in intelligent monitoring and predictive maintenance. 
The task has received increasing attention in recent years, as reflected by the DCASE Task 2 series, which provides public benchmarks for machine sound anomaly detection across different machine types, domains, and operating conditions~\cite{DCASE2020,DCASE2021,DCASE2022,DCASE2023,DCASE2024,DCASE2025}. 
A key challenge of ASD is its normal-only training setting: models are trained using only normal machine sounds and are expected to detect unseen anomalous conditions during evaluation. 
Early ASD systems commonly addressed this setting through reconstruction-based~\cite{IDNN,rushe2019anomaly,jiang2023unsupervised}, or density-estimation~\cite{giri2020unsupervised} approaches, where deviations from normal training data are used as anomaly indicators. 
Subsequent systems further improved performance by exploiting auxiliary machine information, domain-aware training with discriminative classification objectives~\cite{chen2022ss,zhang23j_interspeech,wilkinghoff2024self,jiang2025adaptive}, contrastive learning~\cite{hojjati2022self,guan2023anomalous,zeng2023joint}, and prototype learning~\cite{zeng23b_interspeech,jiang2025adaptive}. 
As recent ASD benchmarks increasingly involve diverse machine types, operating conditions, and source/target domain shifts, learning representations with strong generalization ability has become increasingly important. 
This trend has motivated the use of transferable audio representations learned from large-scale pre-training.

Recently, large-scale pre-trained acoustic representation models have attracted increasing attention for ASD, as they provide transferable representations learned from diverse audio data.
Many recent audio pre-training methods adopt ViT-style architectures~\cite{ViT} over time--frequency patches~\cite{AST,audioMAE,BEATs,EAT,CED}, while Dasheng~\cite{Dasheng} explores a sliding time-patch design for audio representation learning. 
Recent studies have applied such pre-trained models to machine ASD, showing that pre-trained weights can improve robustness when abnormal data are unavailable during training~\cite{han2024exploring,han2025exploring,jiang24c_interspeech}. 
Further improvements have also been reported by using task-specific adaptation or parameter-efficient fine-tuning techniques such as LoRA~\cite{lora,zheng2024improving,han2025exploring}. 
These results suggest the potential of pre-trained acoustic representations for ASD.

However, general-purpose audio pre-training does not explicitly exploit the frequency-structured nature of machine sounds.
Machine sounds often exhibit distinct characteristics across frequency regions, and anomalous cues may appear as localized spectral changes rather than global acoustic patterns~\cite{zhang23fa_interspeech,zeng2022robust,mai2022explaining,zhang2024dual,wang2025lightweight,chen2024enhancing,li2025explainable}.
To capture such frequency-dependent patterns, prior studies have explored attention-based modeling over time--frequency representations~\cite{vaswani2017attention,liu2022anomalous,zhang23fa_interspeech}, shifted-window designs for local modeling across time and frequency~\cite{zhang2024dual,zhang2025multi}, and temporal--spectral fusion strategies for improving ASD performance~\cite{kong2024multi,ma2025estm}.

Motivated by the frequency-structured characteristics of machine sounds, we adopt the ECHO band-splitting architecture~\cite{ECHO} as the basis for developing a more comprehensive cross-band acoustic representation learning framework.
ECHO decomposes the time--frequency representation into multiple sub-bands and learns localized representations within each sub-band, preserving fine-grained spectral information critical for ASD representation learning.
Related band-splitting models, such as FISHER~\cite{FISHER}, also adopt sub-band decomposition, further supporting the effectiveness of this design family.
Nevertheless, ECHO, like other band-splitting models, primarily optimizes sub-band representations independently, and the final representation is typically formed by concatenating the resulting sub-band features. 
As a result, dependencies across sub-bands remain largely implicit, limiting the ability of the model to capture cross-band contextual information during pre-training. Such implicit interaction is insufficient for a pre-trained representation model, where transferable cross-frequency knowledge should be explicitly learned during pre-training.

To address this limitation, we introduce ECHOv2, a frequency-structured pre-trained acoustic representation model with explicit cross-band supervision.
An additional inter-band branch uses global context alignment and masked sub-band reconstruction to encourage information exchange across frequency bands.
The inter-band module is further equipped with multiple summary tokens for structured aggregation, enabling region-aware interaction with controllable granularity along the frequency axis.
This explicit modeling of cross-band dependencies, together with structured aggregation, establishes ECHOv2 as a frequency-structured pre-trained acoustic representation model for ASD.

\begin{table}[t]
\centering
\caption{Comparison between ECHO and ECHOv2 in terms of frequency modeling and pre-training objectives.}
\label{tab:model_diff}
\renewcommand{\arraystretch}{1.15}
\setlength{\tabcolsep}{3pt}
\begin{tabularx}{\columnwidth}{>{\raggedright\arraybackslash}p{0.30\columnwidth}
                                >{\raggedright\arraybackslash}X
                                >{\raggedright\arraybackslash}X}
\hline
\textbf{Aspect} & \textbf{ECHO} & \textbf{ECHOv2} \\
\hline
Cross-frequency modeling
& Implicit
& Explicit via inter-band branch \\ \hline

Training objective
& Intra-band self-distillation
& Two-level self-distillation \\ \hline

Cross-band supervision
& Not explicitly used
& Context alignment and masked sub-band reconstruction \\ \hline

Frequency modeling granularity
& Individual sub-band level
& Structured multi-summary-token level \\ \hline
\end{tabularx}
\end{table}

Alongside model development, evaluating pre-trained representations for ASD in a fair and standardized manner remains challenging, because existing ASD systems often differ in encoders, adaptation strategies, training objectives, and anomaly scoring schemes.
As a result, ASD performance differences may reflect not only the representation capability of a pre-trained model, but also the choice of downstream evaluation pipeline.
Recent benchmark efforts on general audio representations have therefore emphasized standardized and comparable evaluation~\cite{HEAR,SUPERB,X-ARES}. 
HEAR~\cite{HEAR} and SUPERB~\cite{SUPERB} evaluate frozen audio representations using lightweight downstream adaptation modules under standardized benchmark settings. 
X-ARES~\cite{X-ARES} further extends this perspective by incorporating a direct evaluation of raw representations through an unparameterized paradigm.
These benchmarks suggest that representation quality should be examined from complementary perspectives rather than a single evaluation protocol. 

Motivated by these recent benchmarks~\cite{HEAR,SUPERB,X-ARES}, we introduce a unified ASD evaluation benchmark tailored for pre-trained audio backbones. 
This benchmark comprises two complementary protocols: an embedding-based protocol to directly evaluate the intrinsic discriminability of frozen representations, and an adaptation-based protocol to assess the transferability of these representations through a lightweight downstream mapping. 
Together, these protocols provide a standardized and fair platform for comparing pre-trained audio backbones for ASD across multiple datasets and evaluation settings. The differences between ECHOv2 and ECHO are shown in Table~\ref{tab:model_diff}.

Experiments show that ECHO provides a strong band-splitting backbone and outperforms existing strong pre-trained audio baselines under both embedding-based and adaptation-based protocols.
With the proposed two-level distillation strategy, ECHOv2 further improves over ECHO by explicitly incorporating cross-frequency supervision during training.
Ablation studies further indicate that inter-band supervision improves the final band-level representation, and that structured aggregation granularity affects downstream ASD performance.

The main contributions of this work are summarized as follows:
\begin{itemize}
\item We propose ECHOv2, a frequency-structured pre-trained acoustic representation model tailored for machine anomalous sound detection.
\item We introduce a two-level self-distillation strategy with intra-band learning and inter-band supervision, enabling the model to capture both local spectral patterns and global frequency dependencies.
\item We establish a unified evaluation benchmark over DCASE 2020--2025 for standardized, reproducible, and systematic assessment of pre-trained acoustic representations using both frozen representation evaluation and downstream adaptation protocols.
\item Extensive experiments and ablation studies demonstrate the effectiveness of the proposed pre-trained acoustic representation model for industrial acoustic monitoring.
\end{itemize}

\begin{figure}[t]
  \centering
  \includegraphics[width=.9\textwidth]{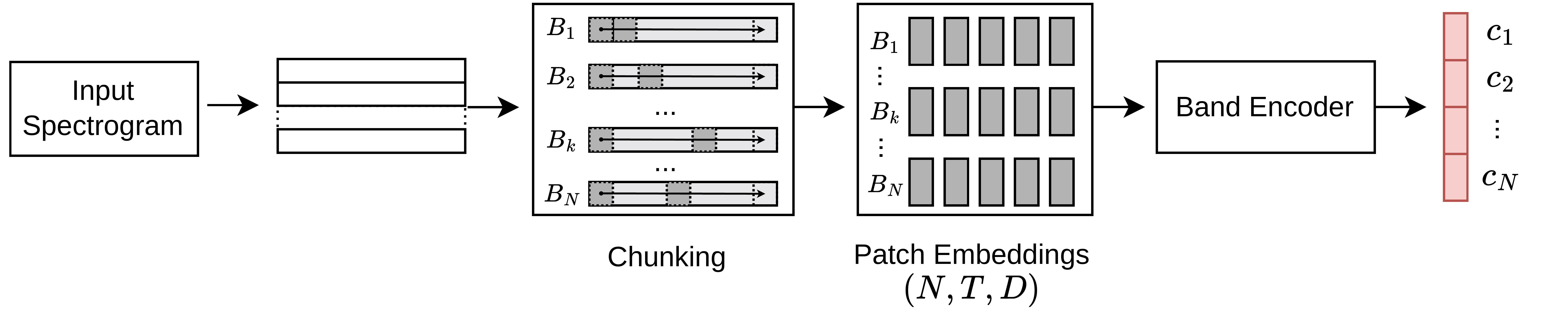}
  \caption{Inference process of the ECHO band-splitting backbone. The input spectrogram is divided into frequency sub-bands, which are encoded by a shared band encoder to produce localized sub-band representations. The final utterance-level representation is obtained by concatenating all sub-band representations.}
  \label{fig:backbone}
\end{figure}

\section{Methodology}

This section first introduces the frequency-splitting acoustic representation backbone adopted in ECHOv2.
We then introduce the proposed cross-band self-distillation framework, which enhances the frequency-splitting representation through explicit inter-band supervision.
The inter-band branch is further extended with structured multi-summary-token aggregation to control the granularity of cross-frequency modeling.
Finally, we describe the training pipeline used to jointly optimize the intra-band and inter-band objectives.

\subsection{Frequency-Splitting Acoustic Representation Backbone}

ECHO~\cite{ECHO} is a band-splitting pre-trained acoustic representation model for ASD.
As shown in Fig.~\ref{fig:backbone}, given an input audio signal, we first convert it into a time--frequency representation. The resulting spectrogram is then split along the frequency axis into $N$ sub-bands:
\begin{equation}
\mathbf{X} \rightarrow \{ \mathbf{X}_k \}_{k=1}^{N}.
\end{equation}

Each sub-band $\mathbf{X}_k$ is further segmented along the time axis into a sequence of non-overlapping local patches. These patches are then projected into embeddings before being fed into the band encoder. Formally, for the $k$-th sub-band, we obtain
\begin{equation}
\mathbf{E}_k = [\mathbf{e}_{k,1}, \mathbf{e}_{k,2}, \ldots, \mathbf{e}_{k,T}],
\end{equation}
where $\mathbf{e}_{k,t}$ denotes the patch embedding at time step $t$ in the $k$-th sub-band, and $T$ is the total number of time steps.

The patch embedding sequence of each sub-band is then processed independently by a shared band encoder:
\begin{equation}
\mathbf{c}_k = f_\theta(\mathbf{E}_k),
\end{equation}
where $\mathbf{c}_k$ denotes the band-level representation of the $k$-th sub-band. The final utterance-level representation is obtained by concatenating the representations from all sub-bands as shown in Fig.~\ref{fig:backbone}:
\begin{equation}
\mathbf{c} = [\mathbf{c}_1, \mathbf{c}_2, \ldots, \mathbf{c}_N].
\end{equation}

This design preserves the frequency-wise decomposition induced by band splitting, while allowing the encoder to model local temporal structure within each sub-band through patch embeddings. The codes and model checkpoints are publicly available~\cite{echo_repo}.

\subsection{Intra-Band Self-Distillation}

\begin{figure}[t]
  \centering
  \includegraphics[width=\columnwidth]{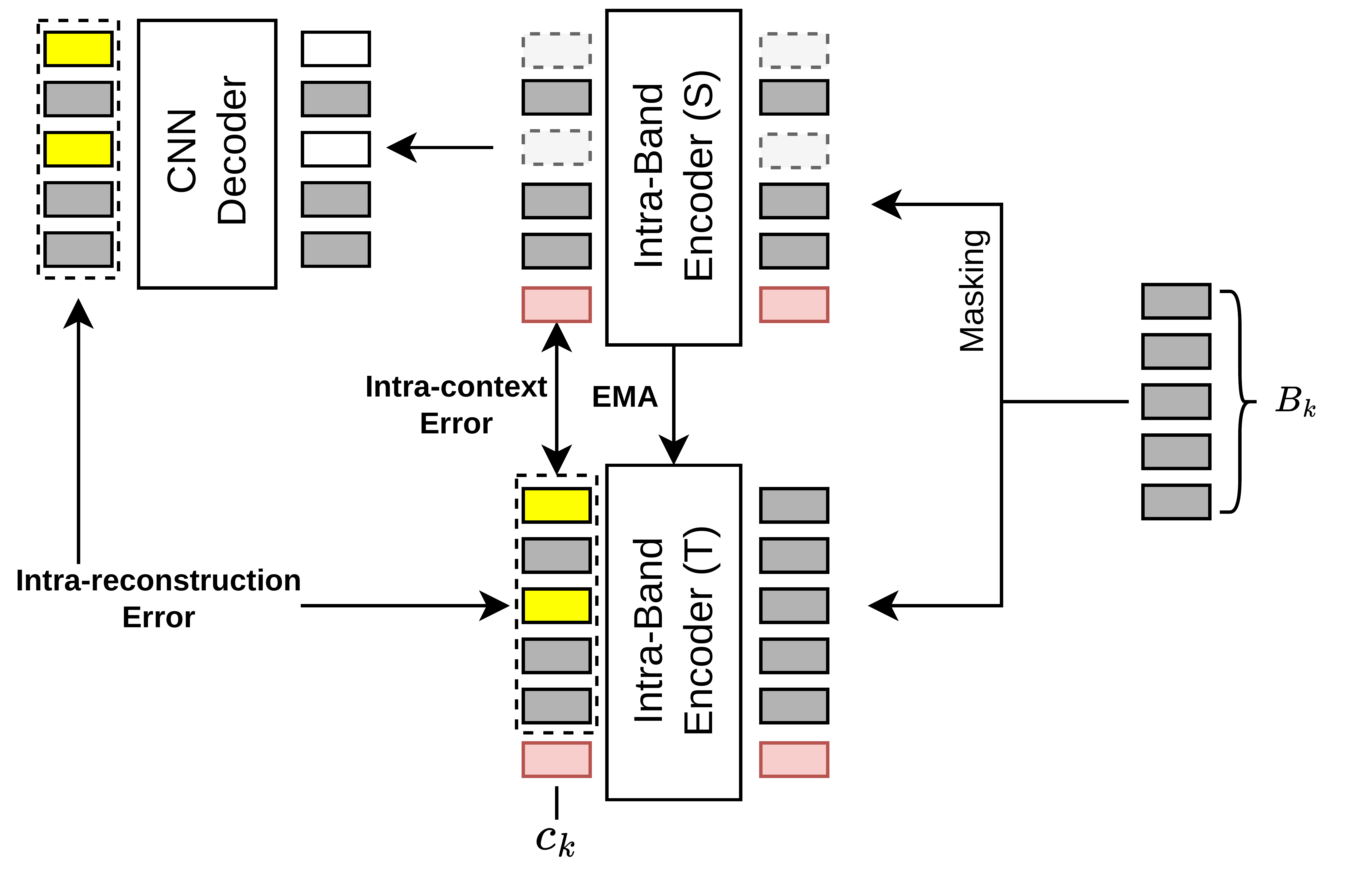} 
  \caption{Intra-band self-distillation for a single sub-band. $B_k$ denotes the $k$-th sub-band, and $\mathbf{c}_k$ denotes the sub-band representation.}
  \label{fig:intra_band} 
\end{figure}

As illustrated in Fig.~\ref{fig:intra_band}, intra-band learning follows the intra-band self-distillation paradigm of the ECHO backbone. Each sub-band is processed independently within a teacher--student self-distillation framework~\cite{EAT}.

For each sub-band, the student encoder receives a masked patch-embedding sequence, while the teacher encoder processes the corresponding unmasked sequence to provide supervision. The student is trained with two complementary objectives: a masked latent reconstruction objective and a representation-level distillation objective.

The reconstruction objective predicts the teacher representations at the masked patch positions using a lightweight convolutional decoder:
\begin{equation}
\mathcal{L}_{\mathrm{rec}}^{(k)} =
\sum_{t \in \mathcal{M}_k}
\left\|
\hat{\mathbf{h}}_{k,t} - \mathbf{h}^{(T)}_{k,t}
\right\|_2^2,
\end{equation}
where $\mathcal{M}_k$ denotes the set of masked positions in the $k$-th sub-band, $\hat{\mathbf{h}}_{k,t}$ is the student prediction for the masked position $t$, and $\mathbf{h}^{(T)}_{k,t}$ is the corresponding teacher encoder embedding.

The distillation objective aligns the student band-level representation with a teacher target:
\begin{equation}
\mathcal{L}_{\mathrm{ctx}}^{\mathrm{intra}} =
\sum_{k}
\left\|
\mathbf{c}_k^{(s)} - \mathbf{t}_k
\right\|_2^2,
\end{equation}
where $\mathbf{c}_k^{(s)}$ is the student representation and $\mathbf{t}_k$ is constructed by aggregating the teacher's intermediate features across layers and then averaging over time:
\begin{equation}
\mathbf{t}_k =
\frac{1}{T}
\sum_{t=1}^{T}
\left(
\frac{1}{L}
\sum_{l=1}^{L}
\mathbf{h}_{k,t}^{(l)}
\right),
\end{equation}
where $\mathbf{h}_{k,t}^{(l)}$ denotes the hidden representation at position $t$ from the $l$-th layer for the $k$-th sub-band. Here, $L$ denotes the total number of layers in the encoder.

As a result, intra-band learning encourages the model to capture both local latent structure and contextual consistency within each frequency band. This branch provides frequency-localized supervision for ECHOv2 and preserves the intra-band modeling ability of ECHO.

\subsection{Inter-Band Self-Distillation}

\begin{figure}[t]
  \centering
  \includegraphics[width=\columnwidth]{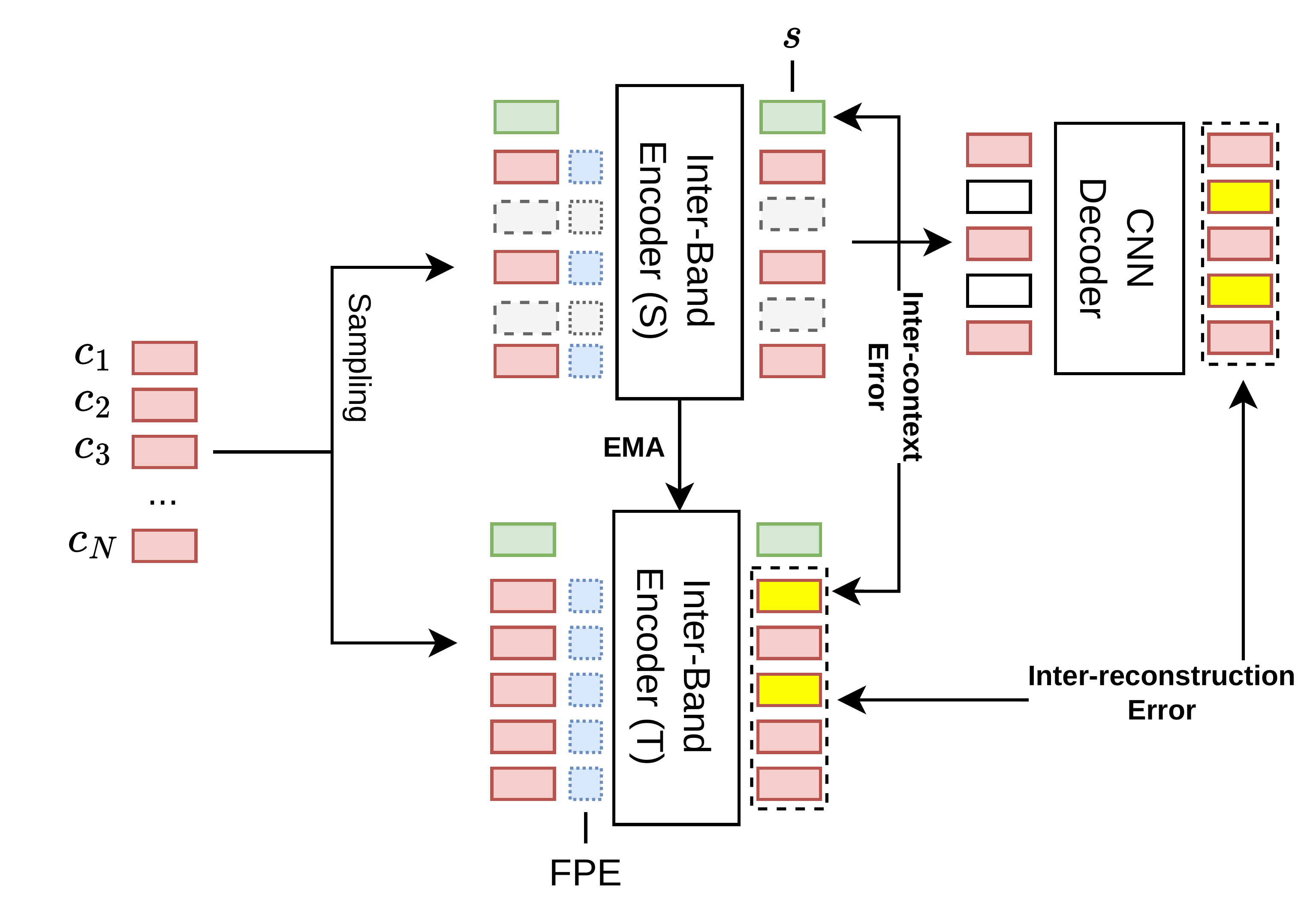} 
  \caption{Inter-band self-distillation over the sequence of sub-band representations. $\mathbf{c}_k$ denotes the $k$-th sub-band representation, and $\mathbf{s}$ denotes the summary token.}
  \label{fig:inter_band} 
\end{figure}

The ECHO intra-band branch provides localized representations, but its supervision is applied independently to each sub-band. As a result, it does not explicitly capture dependencies across frequency bands. To address this limitation, we introduce an inter-band learning mechanism that models cross-band interactions at the representation level.

\paragraph{Cross-band aggregation}

As shown in Fig.~\ref{fig:inter_band}, given band-level representations $\{ \mathbf{c}_k \}_{k=1}^{N}$, we first inject frequency positional information to preserve the relative ordering of sub-bands:
\begin{equation}
\tilde{\mathbf{c}}_k = \mathbf{c}_k + \mathbf{p}_k,
\end{equation}
where $\mathbf{p}_k$ is the frequency positional embedding~(FPE) of the $k$-th band. Specifically,
\begin{equation}
\mathbf{p}_k = \mathrm{PE}(r_k),
\qquad
r_k = \frac{f_k - f_{\min}}{f_{\max} - f_{\min}} \in (0,1),
\end{equation}
where $f_k$ denotes the center frequency of the $k$-th sub-band, and $\mathrm{PE}(\cdot)$ is the sinusoidal positional encoding function. In this way, the inter-band encoder is informed not only by the content of each band representation, but also by its relative location on the frequency axis.

The position-aware band tokens are then processed by an inter-band encoder alongside a summary token $\mathbf{s}$:
\begin{equation}
[\mathbf{s}', \mathbf{u}_1, \ldots, \mathbf{u}_N]
=
g_\theta([\mathbf{s}, \tilde{\mathbf{c}}_1, \ldots, \tilde{\mathbf{c}}_N]),
\end{equation}
where $\mathbf{s}'$ denotes the updated summary token and $\mathbf{u}_k$ denotes the updated representation of the $k$-th band. The summary token aggregates global cross-band context, while the updated band tokens retain band-specific information after interacting with other frequency regions.

\begin{figure}[t]
  \centering
  \includegraphics[width=.6\textwidth]{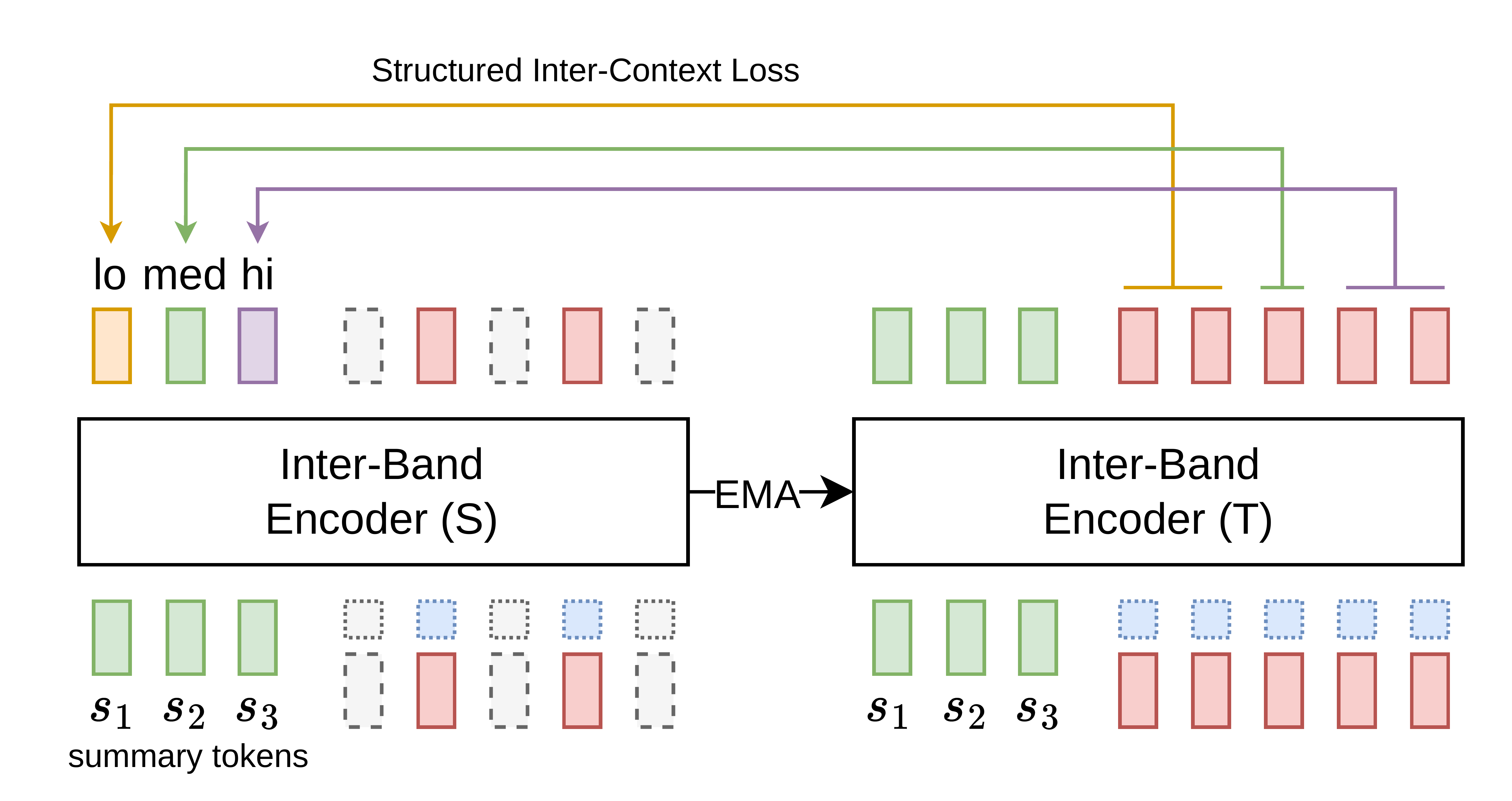} 
  \caption{Structured inter-band learning with multiple summary tokens. $s_k$ denotes the summary token.}
  \label{fig:structured} 
\end{figure}

\paragraph{Inter-band objectives}

Inter-band learning adopts the same teacher--student self-distillation paradigm as intra-band learning, with a context-matching objective and a reconstruction objective defined at the band-representation level. The key difference is that the modeling unit is now the band-level representation sequence rather than the patch sequence within each sub-band. Accordingly, the supervision is imposed in the inter-band representation space to encourage cross-band dependency modeling.

\textit{1) Inter-context loss:}
The student summary token $\mathbf{s}'$ is used as a global representation of the band sequence and is aligned with a teacher target constructed by aggregating the teacher's intermediate inter-band representations across layers and bands:
\begin{equation}
\mathbf{t}
=
\frac{1}{N}
\sum_{k=1}^{N}
\left(
\frac{1}{L}
\sum_{l=1}^{L}
\mathbf{h}^{(l)}_{k}
\right),
\end{equation}
where $\mathbf{h}^{(l)}_{k}$ denotes the hidden representation of the $k$-th band at the $l$-th layer of the teacher inter-band encoder. The corresponding loss is
\begin{equation}
\mathcal{L}^{\mathrm{inter}}_{\mathrm{ctx}} =
\left\|
\mathbf{s}' - \mathbf{t}
\right\|_2^2.
\end{equation}

\textit{2) Inter-reconstruction loss:}
In parallel, missing band representations are reconstructed in the latent space. Specifically, a lightweight decoder predicts the teacher representation at each masked band position:
\begin{equation}
\mathcal{L}_{\mathrm{rec}}^{\mathrm{inter}} =
\sum_{k \in \mathcal{M}}
\left\|
\hat{\mathbf{u}}_k - \mathbf{u}^{(T)}_k
\right\|_2^2,
\end{equation}
where $\hat{\mathbf{u}}_k$ is the prediction for the $k$-th masked band and $\mathbf{u}^{(T)}_k$ is the corresponding teacher representation.

As a result, inter-band learning encourages the model to capture both holistic cross-band information and dependency structure among band representations, aiming to complement intra-band learning by explicitly modeling interactions across different frequency regions.

\subsection{Structured Inter-Band Aggregation}

The inter-band formulation above relies on a single summary token, which performs global aggregation over all frequency bands. Although this design is effective, a single token may be insufficient to represent heterogeneous spectral patterns distributed across different frequency regions.

To provide a more structured form of cross-band aggregation, we extend the inter-band module by introducing multiple summary tokens. As illustrated in Fig.~\ref{fig:structured}, instead of forcing all cross-band information into a single global token, the model is allowed to maintain several aggregation pathways, each of which can attend to different subsets of band representations.

Concretely, let $\{\mathbf{s}_m\}_{m=1}^{M}$ denote $M$ learnable summary tokens. The inter-band encoder processes the augmented token sequence as
\begin{equation}
[\mathbf{s}_1', \ldots, \mathbf{s}_M', \mathbf{u}_1, \ldots, \mathbf{u}_N]
=
g_\theta([\mathbf{s}_1, \ldots, \mathbf{s}_M, \tilde{\mathbf{c}}_1, \ldots, \tilde{\mathbf{c}}_N]),
\end{equation}
where $\mathbf{s}_m'$ denotes the updated $m$-th summary token and $\mathbf{u}_k$ denotes the updated representation of the $k$-th band.

Each summary token is encouraged to serve as a learnable aggregation unit for a frequency-ordered group of bands. In this way, inter-band learning is reformulated from a single global aggregation path to a multi-granularity structured aggregation scheme.

The inter-context objective is generalized accordingly:
\begin{equation}
\mathcal{L}^{\mathrm{inter}}_{\mathrm{ctx}}
=
\frac{1}{M}\sum_{m=1}^{M}
\left\|
\mathbf{s}_m' - \mathbf{t}_m
\right\|_2^2,
\end{equation}
where $\mathbf{t}_m$ denotes the teacher target associated with the $m$-th summary token. Specifically, each $\mathbf{t}_m$ is obtained by aggregating the teacher inter-band representations over a frequency-ordered subset of bands:
\begin{equation}
\mathbf{t}_m =
\frac{1}{|\mathcal{G}_m|}
\sum_{k \in \mathcal{G}_m}
\left(
\frac{1}{L}
\sum_{l=1}^{L}
\mathbf{h}^{(l)}_{k}
\right),
\end{equation}
where $\mathcal{G}_m$ denotes the set of bands associated with the $m$-th summary token. Here, the full band sequence is partitioned into $M$ contiguous groups along the frequency axis, and $\mathcal{G}_m$ corresponds to the $m$-th such group.

The number of summary tokens $M$ controls the granularity of cross-band supervision. When $M=1$, the formulation reduces to the single-token global inter-band aggregation. Increasing $M$ enables the model to represent cross-band dependencies at a finer granularity, which is useful when spectral patterns from different frequency regions exhibit distinct characteristics.

This design improves the flexibility of cross-band supervision without changing the underlying ECHO backbone.

\begin{figure}[t]
  \centering
  \includegraphics[width=\columnwidth]{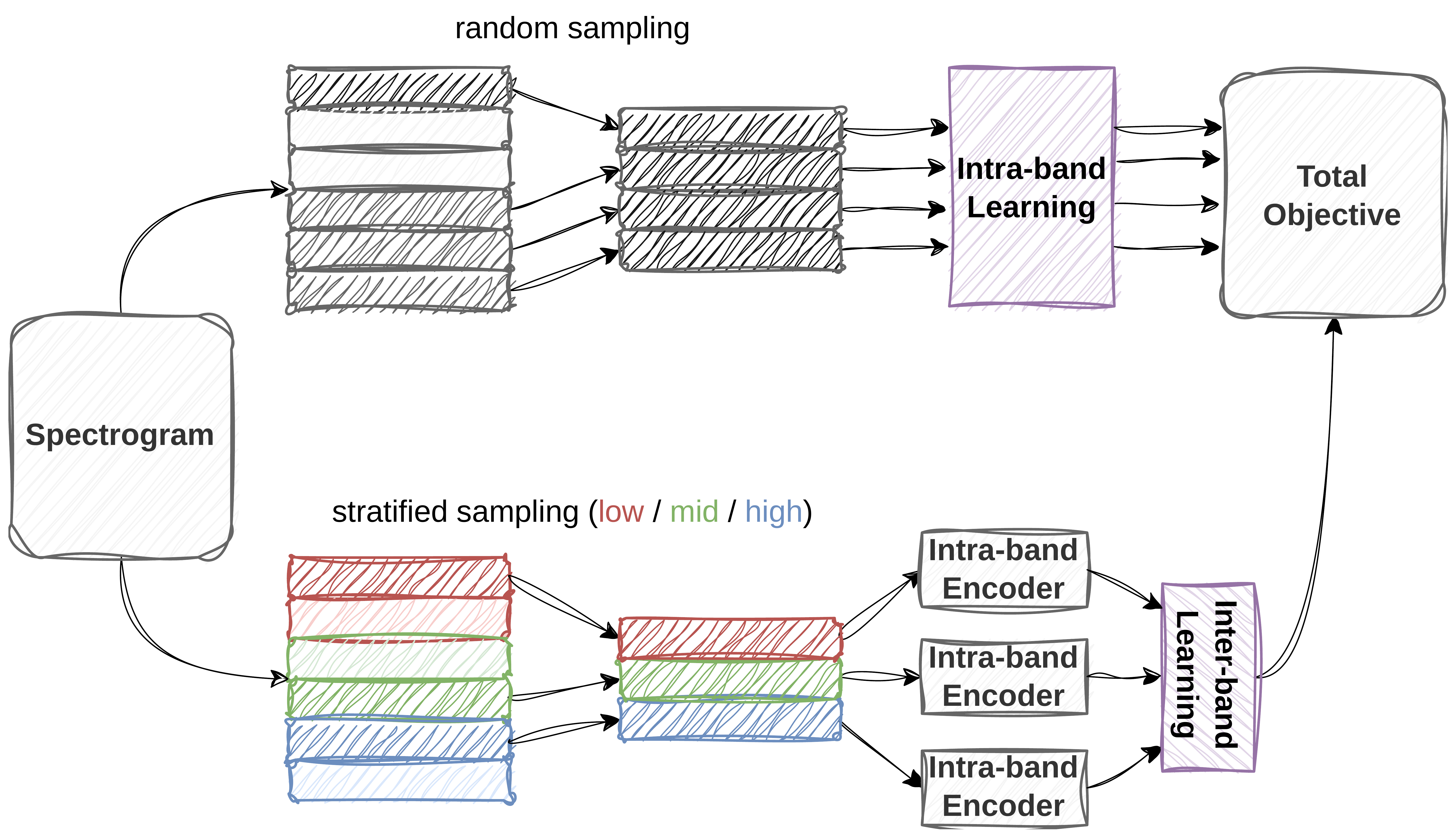}
  \caption{Band-level, student-side view of the proposed training pipeline. Starting from an input spectrogram, the student branch forms two parallel pathways: intra-band learning with random sub-band sampling, and inter-band learning with stratified frequency sampling over low-, medium-, and high-frequency regions. The sampled sub-bands are processed by the intra-band encoder, and the two branches are jointly optimized under a unified objective. Teacher supervision is omitted for clarity.}
  \label{fig:training_pipeline}
\end{figure}

\subsection{Training Pipeline}

Fig.~\ref{fig:training_pipeline} illustrates the overall training process from a band-level, student-side perspective. Given an input spectrogram, the student branch constructs two parallel pathways with different sub-band sampling strategies, corresponding to intra-band and inter-band learning.

For intra-band learning, sub-bands are randomly sampled from the full set of available sub-bands, and each sampled sub-band is processed independently by the intra-band branch. This pathway follows the ECHO intra-band training pipeline and focuses on within-band representation learning through reconstruction and contextual alignment at the individual band level.

For inter-band learning, the student branch instead adopts stratified sampling to construct a band sequence for cross-band modeling. Specifically, the full set of sub-bands is first divided, in ascending frequency order, into three contiguous frequency groups corresponding to low-, medium-, and high-frequency regions. The same number of sub-bands is then sampled from each group, so that the resulting sequence maintains balanced spectral coverage. These sampled sub-bands are first encoded into band-level representations by the intra-band encoder, and the resulting representations are then fed into the inter-band module for cross-band aggregation and supervision.

The teacher branch is not shown in Fig.~\ref{fig:training_pipeline} for simplicity. In the intra-band branch, the teacher and student follow the same band-wise pathway, except that the student receives masked inputs and the teacher processes unmasked inputs. In contrast, in the inter-band branch, the stratified sampling is applied only to the student input, whereas the teacher continues to process the complete band set and provides supervisory targets derived from the full-spectrum band sequence.

The two pathways are optimized jointly under a unified objective:
\begin{equation}
\mathcal{L}_{\mathrm{total}}
=
\lambda_1\cdot\mathcal{L}^{\mathrm{intra}}_{\mathrm{rec}}
+
\lambda_2\cdot\mathcal{L}^{\mathrm{intra}}_{\mathrm{ctx}}
+
\lambda_3\cdot\mathcal{L}^{\mathrm{inter}}_{\mathrm{rec}}
+
\lambda_4\cdot\mathcal{L}^{\mathrm{inter}}_{\mathrm{ctx}},
\end{equation}
where $\mathcal{L}^{\mathrm{intra}}_{\mathrm{rec}}$ and $\mathcal{L}^{\mathrm{intra}}_{\mathrm{ctx}}$ denote the intra-band reconstruction and contextual alignment losses, respectively, and $\mathcal{L}^{\mathrm{inter}}_{\mathrm{rec}}$ and $\mathcal{L}^{\mathrm{inter}}_{\mathrm{ctx}}$ denote their inter-band counterparts. This joint formulation couples intra-band and inter-band learning within a single training strategy, allowing the model to retain band-wise modeling capacity while explicitly incorporating cross-band dependencies.

\begin{figure}[t]
  \centering
  \includegraphics[width=.85\textwidth]{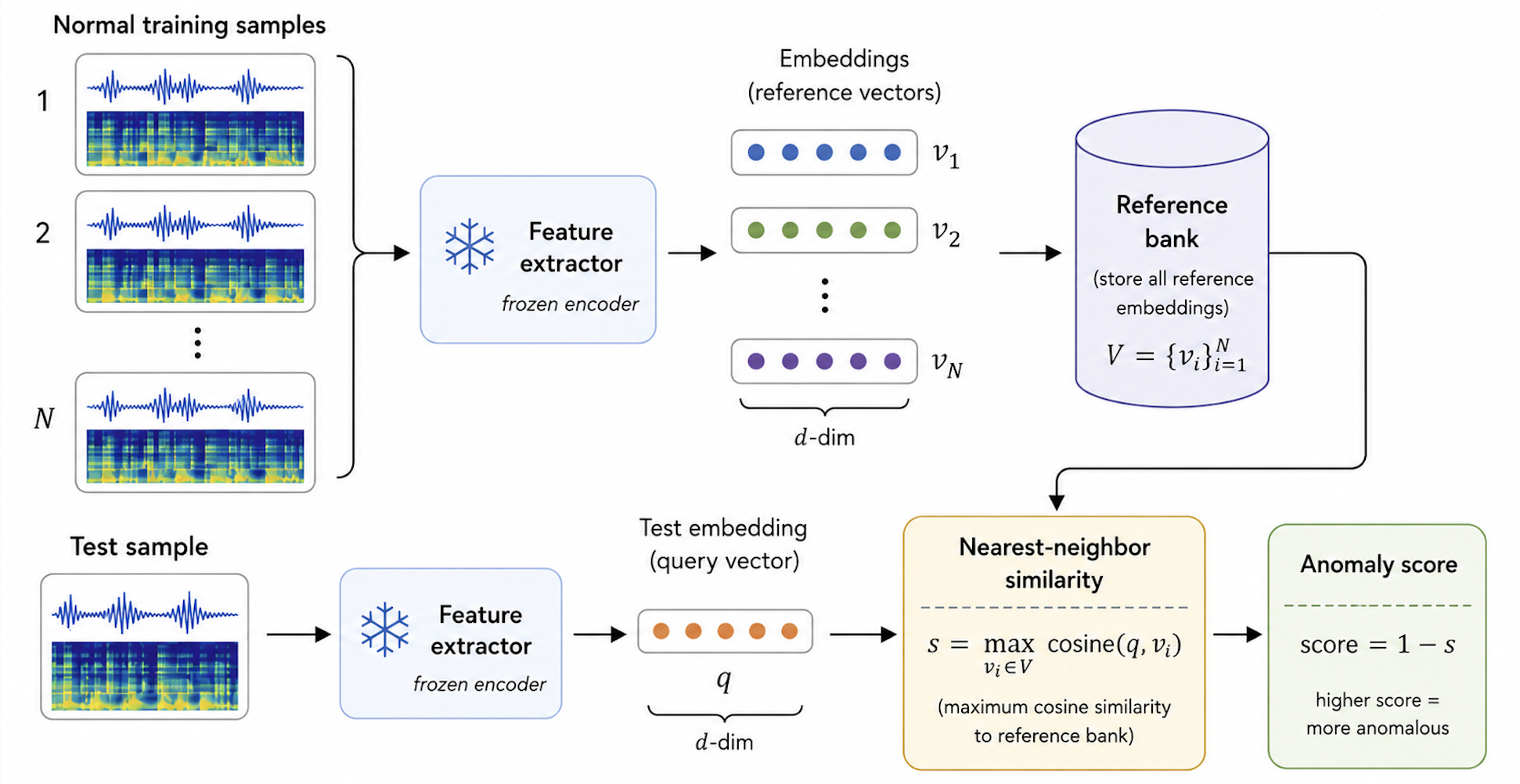}
  \caption{Embedding-based anomaly scoring protocol. Normal training samples are first encoded into reference embeddings to construct a reference bank. For each test sample, a frozen encoder extracts a query embedding, which is compared with the reference bank using nearest-neighbor cosine similarity. The anomaly score is computed as $1-s$, where $s$ denotes the maximum cosine similarity to the reference bank.}
  \label{fig:embedding_protocol}
\end{figure}

\begin{table}[t]
\centering
\caption{Year-specific evaluation settings in the unified ASD benchmark. MTYPE, MID, SEC, and DOM denote machine type, machine ID, section, and domain, respectively.}
\label{tab:year_specific_protocol}
\begin{tabularx}{\textwidth}{
l
>{\raggedright\arraybackslash}X
>{\raggedright\arraybackslash}X
>{\raggedright\arraybackslash}X
}
\toprule
\textbf{Year} & \textbf{Task characteristic} & \textbf{Scoring granularity} & \textbf{Final aggregation} \\
\midrule
2020 
& ID-aware ASD 
& (MTYPE,MID) for AUC and pAUC 
& Arithmetic mean \\

2021 
& Domain-shifted ASD 
& (MTYPE,SEC,DOM) for AUC and pAUC 
& Harmonic mean \\

2022--2025 
& Section/domain-aware ASD 
& (MTYPE,SEC,DOM) for AUC; (MTYPE,SEC) for pAUC 
& Harmonic mean \\
\bottomrule
\end{tabularx}
\end{table}

\section{Evaluation of Pre-trained Acoustic Representations for Machine Anomalous Sound Detection}

\noindent
Evaluating pre-trained acoustic representations for ASD requires a standardized protocol that separates representation quality from downstream system design.
Existing ASD systems often differ in encoders, adaptation strategies, training objectives, and anomaly scoring schemes, making direct comparison among pre-trained backbones difficult.
To address this issue, we establish a unified ASD evaluation benchmark over DCASE 2020--2025 for comparing pre-trained audio backbones under consistent evaluation settings.

The benchmark is designed to assess pre-trained audio backbones from two complementary perspectives.
The embedding-based protocol directly evaluates the intrinsic discriminability of frozen representations without introducing task-specific adaptation modules.
The adaptation-based protocol evaluates representation transferability by applying a lightweight downstream projection trained on normal data before anomaly scoring.
Together, these protocols provide a unified platform for comparing pre-trained audio backbones under consistent ASD evaluation settings.

\subsection{DCASE Datasets}

\noindent
Our experiments are conducted exclusively on the DCASE Task 2 ASD datasets~\cite{DCASE2020,DCASE2021,DCASE2022,DCASE2023,DCASE2024,DCASE2025}, which are generated from other popular machine sound datasets. 
Specifically, the DCASE 2020 dataset is generated from MIMII~\cite{MIMII} and ToyADMOS~\cite{ToyADMOS}; the DCASE 2021 dataset is from MIMII Due~\cite{MIMII_DUE} and ToyADMOS2~\cite{ToyADMOS2}; the DCASE 2022 ASD dataset is from MIMII DG~\cite{MIMII_DG} and ToyADMOS2~\cite{ToyADMOS2}; the DCASE 2023 ASD dataset is from MIMII DG~\cite{MIMII_DG} and ToyADMOS2+~\cite{ToyADMOS2+}; the DCASE 2024 ASD dataset is from MIMII DG~\cite{MIMII_DG} and ToyADMOS2\#~\cite{ToyADMOS2++}, and finally the DCASE 2025 ASD dataset is from MIMII DG~\cite{MIMII_DG}, ToyADMOS2025~\cite{ToyADMOS2025} and IMAD-DS~\cite{IMAD-DS}. 

These six benchmarks all target machine condition monitoring, while reflecting the evolution of the task setting from conventional unsupervised ASD to domain-shifted or domain-generalization scenarios and, more recently, first-shot unsupervised ASD. 
Depending on the year, each benchmark contains recordings from various numbers of machine types. 
The input audio consists mainly of short single-channel clips of approximately 10 seconds, typically containing both target-machine sounds and environmental noise. 
For all six benchmarks, we use the official data splits provided by the corresponding challenges. 

\subsection{Year-Specific DCASE Evaluation Protocols}
\label{subsec:yearProtocols}

\noindent
Table~\ref{tab:year_specific_protocol} summarizes the year-specific evaluation definitions adopted in the unified ASD benchmark.
Across all years, the final submission format consists of sample-level anomaly scores, while the official performance is evaluated using AUC and pAUC. 
Here, pAUC denotes the partial area under the ROC curve computed over the false-positive-rate range $[0, 0.1]$.

For DCASE 2020, both AUC and pAUC are computed at the machine-ID level within each machine type, and the final yearly score is obtained by arithmetic averaging over all reported values. 
For DCASE 2021, both AUC and pAUC are computed at the granularity of machine type, section, and domain, and the final yearly score is given by the harmonic mean over all reported values. 
For DCASE 2022--2025, AUC is computed at the granularity of machine type, section, and domain, whereas pAUC is computed at the granularity of machine type and section. 
The final yearly score for these benchmarks is again obtained by the harmonic mean.

In our implementation, these official evaluation definitions are applied after anomaly scores are generated by the corresponding benchmark protocol.

\subsection{Frozen Representation Evaluation Protocol}

\noindent
The frozen representation protocol evaluates frozen representations directly in the feature space.
Fig.~\ref{fig:embedding_protocol} illustrates the anomaly scoring process.
For each benchmark year, embeddings extracted from normal training samples are stored as reference features.
Each test sample is assigned an anomaly score by comparing its embedding with the corresponding reference bank.
All embeddings are $\ell_2$-normalized, and the anomaly score is computed as the nearest-neighbor cosine distance.

The construction of the reference bank follows the year-specific metadata used for matching and scoring, as summarized in Table~\ref{tab:year_specific_protocol}.
For DCASE 2020, reference matching is performed using machine type and machine ID.
For DCASE 2021--2025, reference matching is performed using machine type, section, and domain information.
Final performance is computed using the official AUC and pAUC definitions of each year.

\subsection{Downstream Adaptation Evaluation Protocol}

\noindent
The downstream adaptation protocol evaluates whether frozen representations can be effectively transferred to a downstream ASD setting.
For each benchmark year, we train a lightweight adaptation module on top of frozen embeddings using only normal training data with auxiliary labels derived from the available metadata.
The module consists of a linear projection layer followed by a linear classifier.
During adaptation, the pre-trained encoder remains frozen, and only the lightweight module is optimized.

Auxiliary labels are constructed from the metadata available in each DCASE year, such as machine type, machine ID, section, domain, and operating-condition attributes.
After training, the classifier is discarded, and only the learned projection layer is retained to transform embeddings into an adapted feature space.
Anomaly scoring is then performed using the same memory-bank protocol illustrated in Fig.~\ref{fig:embedding_protocol}.

\begin{sidewaystable}[p]
\centering
\Large
\caption{Main ASD results (\%) under the embedding-based evaluation protocol on DCASE 2020--2025. 
For DCASE 2020, the yearly score is the arithmetic mean of the development and evaluation results. 
For DCASE 2021--2025, the yearly score is the harmonic mean. 
``DEV'' and ``EVAL'' denote development and evaluation scores under the year-specific protocols in Section~\ref{subsec:yearProtocols}. 
``Overall'' denotes the arithmetic mean over the six yearly scores. 
Best and second-best results are marked in bold and underline, respectively.}
\label{tab:asd_main_raw}
\resizebox{\textheight}{!}{
\begin{tabular}{cc|ccc|ccc|ccc|ccc|ccc|ccc|c}
\toprule
\multirow{2}{*}{\textbf{Model}} & \multirow{2}{*}{\textbf{Version}} 
& \multicolumn{3}{c|}{\textbf{2020}} 
& \multicolumn{3}{c|}{\textbf{2021}} 
& \multicolumn{3}{c|}{\textbf{2022}} 
& \multicolumn{3}{c|}{\textbf{2023}} 
& \multicolumn{3}{c|}{\textbf{2024}} 
& \multicolumn{3}{c|}{\textbf{2025}} 
& \multirow{2}{*}{\textbf{Overall}} \\
& 
& \textbf{DEV} & \textbf{EVAL} & \textbf{Mean} 
& \textbf{DEV} & \textbf{EVAL} & \textbf{HMean} 
& \textbf{DEV} & \textbf{EVAL} & \textbf{HMean} 
& \textbf{DEV} & \textbf{EVAL} & \textbf{HMean} 
& \textbf{DEV} & \textbf{EVAL} & \textbf{HMean} 
& \textbf{DEV} & \textbf{EVAL} & \textbf{HMean} & \\
\midrule
\multirow{1}{*}{BEATs~\cite{BEATs}} 
& base
& \textbf{73.71} & \underline{74.98} & \textbf{74.26}
& \textbf{63.43} & {\textbf{59.00}} & \textbf{61.31}
& 62.90 & 55.50 & 58.97
& 60.44 & 65.54 & 62.89
& 55.96 & 55.83 & 55.89
& 59.26 & 56.65 & 57.84
& 61.86 \\
\midrule
\multirow{4}{*}{CED~\cite{CED}} 
& base
& 67.40 & 68.20 & 67.75
& 56.87 & 56.47 & 56.67
& 59.50 & 55.18 & 57.26
& 59.79 & 61.94 & 60.84
& \underline{57.91} & 57.77 & 57.83
& 59.82 & 55.99 & 57.72
& 59.68 \\
& mini
& 66.86 & 68.54 & 67.59
& 56.50 & 56.20 & 56.35
& 59.53 & 54.73 & 57.03
& 59.08 & 60.63 & 59.85
& 57.66 & 55.44 & 56.39
& 58.54 & 56.48 & 57.43
& 59.11 \\
& small
& 67.47 & 67.98 & 67.69
& 57.10 & 56.21 & 56.66
& 59.42 & 54.39 & 56.79
& 59.07 & 61.04 & 60.04
& \textbf{58.00} & 56.67 & 57.24
& 59.92 & 56.22 & 57.89
& 59.39 \\
& tiny
& 67.14 & 67.30 & 67.21
& 56.25 & 56.09 & 56.17
& 58.96 & 54.73 & 56.77
& 58.48 & 60.79 & 59.61
& 57.82 & 54.98 & 56.19
& 59.13 & 56.72 & 57.82
& 58.96 \\
\midrule
\multirow{3}{*}{Dasheng~\cite{Dasheng}} 
& base
& 69.25 & 69.03 & 69.15
& 58.00 & 56.57 & 57.27
& 60.86 & 55.16 & 57.87
& 59.92 & 61.50 & 60.70
& 56.90 & 58.36 & 57.71
& 58.14 & 56.06 & 57.01
& 59.95 \\
& 0.6b
& 68.22 & 68.13 & 68.18
& 57.30 & 56.22 & 56.76
& 59.21 & 54.23 & 56.61
& 58.27 & 61.72 & 59.94
& 55.58 & 57.69 & 56.75
& 57.77 & 55.85 & 56.73
& 59.16 \\
& 1.2b
& 69.58 & 69.34 & 69.48
& 58.21 & 55.95 & 57.06
& 60.36 & 54.53 & 57.29
& 59.35 & 63.23 & 61.23
& 55.84 & 58.17 & 57.13
& 58.89 & 55.65 & 57.12
& 59.88 \\
\midrule
\multirow{2}{*}{EAT~\cite{EAT}} 
& base
& 70.49 & 74.23 & 72.13
& 58.01 & 57.57 & 57.79
& 61.11 & 56.24 & 58.57
& 59.50 & 59.89 & 59.69
& 56.99 & 57.22 & 57.12
& \underline{60.81} & 58.86 & \underline{59.75}
& 60.84 \\
& large
& \underline{72.43} & \textbf{75.86} & \underline{73.94}
& 58.57 & 56.41 & 57.47
& 62.31 & 55.20 & 58.54
& 60.20 & 63.20 & 61.66
& 57.66 & 58.07 & \underline{57.89}
& \textbf{62.33} & 58.12 & \textbf{60.01}
& 61.58 \\
\midrule
\multirow{3}{*}{FISHER~\cite{FISHER}} 
& mini
& 68.42 & 71.97 & 69.98
& 60.06 & 56.80 & 58.39
& 60.76 & 55.32 & 57.91
& 59.10 & 61.65 & 60.35
& 54.91 & 56.72 & 55.91
& 59.76 & 54.61 & 56.90
& 59.91 \\
& small
& 69.10 & 72.38 & 70.54
& 60.34 & 58.70 & 59.51
& \underline{64.05} & 56.06 & 59.79
& 59.80 & 64.01 & 61.83
& 55.20 & 56.02 & 55.66
& 59.79 & 57.75 & 58.68
& 61.00 \\
& tiny
& 68.96 & 72.78 & 70.64
& 60.14 & 56.96 & 58.51
& 59.22 & 55.14 & 57.11
& 58.10 & 58.83 & 58.46
& 53.55 & 56.82 & 55.34
& 58.62 & 56.90 & 57.69
& 59.62 \\
\midrule
\multirow{2}{*}{ECHO~\cite{ECHO}} 
& small
& 70.96 & 73.84 & 72.23
& 61.77 & 58.70 & 60.20
& 63.98 & \underline{56.42} & \underline{59.96}
& \textbf{61.40} & \underline{66.21} & \underline{63.71}
& 55.63 & \underline{59.73} & 57.86
& 58.97 & 58.47 & 58.70
& \underline{62.11} \\
& tiny
& 68.69 & 72.00 & 70.14
& 60.01 & 58.03 & 59.01
& 63.62 & 56.34 & 59.76
& \underline{61.33} & \textbf{66.36} & \textbf{63.75}
& 55.52 & 58.05 & 56.91
& 57.07 & \textbf{59.62} & 58.40
& 61.33 \\
\midrule
\multirow{1}{*}{ECHOv2} 
& small
& 71.69 & 74.59 & 72.96
& \underline{61.78} & {\textbf{59.00}} & \underline{60.36}
& \textbf{64.23} & \textbf{57.13} & \textbf{60.48}
& 61.13 & 66.17 & 63.55
& 55.71 & \textbf{61.45} & \textbf{58.80}
& 59.77 & \underline{59.49} & 59.62
& \textbf{62.63} \\
\bottomrule
\end{tabular}
}
\end{sidewaystable}

\section{Experimental Results}

\subsection{Experimental Setup}

\subsubsection{Baselines}

We compare ECHOv2 with several representative pre-trained acoustic representation models, including BEATs~\cite{BEATs}, CED~\cite{CED}, EAT~\cite{EAT}, Dasheng~\cite{Dasheng}, and FISHER~\cite{FISHER}. 
These baselines cover both general-purpose acoustic encoders and band-splitting-based models. 
All compared models are publicly available and are built upon Transformer- or ViT-style backbones pre-trained on large-scale open-source audio corpora from different domains, including general audio datasets, such as AudioSet~\cite{audioset}, VGGSound~\cite{vggsound}, ACAV100M~\cite{acav100m}, and Freesound~\cite{freesound}, and music datasets, such as MTG-Jamendo~\cite{mtgjamendo}, and Music4All~\cite{music4all}.
ECHO is included as the direct band-splitting baseline, allowing us to isolate the effect of the proposed inter-band supervision and structured aggregation in ECHOv2.
 
\subsubsection{Implementation Details}
ECHOv2 uses the ECHO backbone, and its default downstream representation follows the ECHO setting by concatenating band-level features from all sub-bands.
The input is a spectrogram extracted from normalized raw audio using a 25 ms window and a 10 ms hop, and the sub-band width is fixed to 32.
For the inter-band branch, the encoder consists of two Transformer layers with four attention heads, and the default number of summary tokens is set to three.

We train the model with AdamW~\cite{adamw} using a two-stage schedule. 
In the first stage, the backbone is optimized with only the intra-band objective for 400k steps, where $\lambda_1=\lambda_2=0.5$. 
In the second stage, the model is jointly optimized with both intra-band and inter-band learning enabled for another 100k steps, where $\lambda$ coefficients of the training objectives are equally divided into $0.25$. 
For both stages, the learning rate is linearly warmed up to $1\times10^{-4}$ and then cosine-decayed to $1\times10^{-5}$. 
The batch size is 256 and the weight decay is 0.05. 
Pre-training follows the ECHO backbone setup and uses public audio corpora, mainly including AudioSet~\cite{audioset}, MTG-Jamendo~\cite{mtgjamendo}, and Freesound audio data from WavCaps~\cite{wavcaps}. 
Compared with several baselines that additionally use datasets such as VGGSound~\cite{vggsound}, ACAV100M~\cite{acav100m}, or Music4All~\cite{music4all}, ECHOv2 is trained without these extra corpora, which makes the comparison conservative with respect to pre-training data scale.
All experiments were performed on two computing nodes with a total of eight K500SM\_AI 64 GB accelerator cards.

For the adaptation-based ASD evaluation, all models use the same lightweight adaptation setting. 
A linear projection layer with output dimension 256 and a linear classifier are trained on top of frozen embeddings using AdamW for 10 epochs, with a learning rate of $1\times10^{-3}$, weight decay of $1\times10^{-4}$, and batch size of 64. 
After training, only the learned projection layer is retained for subsequent ASD evaluation.

\begin{sidewaystable}[p]
\centering
\Large
\caption{Main ASD results (\%) under the adaptation-based evaluation protocol on DCASE 2020--2025. 
For DCASE 2020, the yearly score is the arithmetic mean of the development and evaluation results. 
For DCASE 2021--2025, the yearly score is the harmonic mean. 
``DEV'' and ``EVAL'' denote development and evaluation scores under the year-specific protocols in Section~\ref{subsec:yearProtocols}. 
``Overall'' denotes the arithmetic mean over the six yearly scores. 
Best and second-best results are marked in bold and underline, respectively.}
\label{tab:asd_main_adapt}
\resizebox{\textheight}{!}{
\begin{tabular}{cc|ccc|ccc|ccc|ccc|ccc|ccc|c}
\toprule
\multirow{2}{*}{\textbf{Model}} & \multirow{2}{*}{\textbf{Version}} 
& \multicolumn{3}{c|}{\textbf{2020}} 
& \multicolumn{3}{c|}{\textbf{2021}} 
& \multicolumn{3}{c|}{\textbf{2022}} 
& \multicolumn{3}{c|}{\textbf{2023}} 
& \multicolumn{3}{c|}{\textbf{2024}} 
& \multicolumn{3}{c|}{\textbf{2025}} 
& \multirow{2}{*}{\textbf{Overall}} \\
& 
& \textbf{DEV} & \textbf{EVAL} & \textbf{Mean} 
& \textbf{DEV} & \textbf{EVAL} & \textbf{HMean} 
& \textbf{DEV} & \textbf{EVAL} & \textbf{HMean} 
& \textbf{DEV} & \textbf{EVAL} & \textbf{HMean} 
& \textbf{DEV} & \textbf{EVAL} & \textbf{HMean} 
& \textbf{DEV} & \textbf{EVAL} & \textbf{HMean} & \\
\midrule
\multirow{1}{*}{BEATs~\cite{BEATs}}
& base
& \underline{77.41} & \textbf{82.75} & \textbf{79.75}
& 61.23 & 58.84 & 60.01
& 63.84 & 56.90 & 60.17
& 59.93 & 61.93 & 60.91
& \textbf{59.25} & 57.55 & 58.28
& 59.53 & 55.23 & 57.15
& 62.71 \\
\midrule
\multirow{4}{*}{CED~\cite{CED}}
& base
& 72.91 & 75.12 & 73.88
& 59.17 & 56.43 & 57.77
& 62.06 & 56.52 & 59.16
& 58.88 & 60.46 & 59.66
& 58.69 & 56.15 & 57.24
& 59.14 & 56.51 & 57.71
& 60.90 \\
& mini
& 71.79 & 74.81 & 73.11
& 58.32 & 56.28 & 57.28
& 61.33 & 56.69 & 58.92
& 57.80 & 59.88 & 58.82
& 57.62 & 55.53 & 56.43
& 58.00 & 55.56 & 56.68
& 60.21 \\
& small
& 72.91 & 74.91 & 73.79
& 59.28 & 56.37 & 57.79
& 61.91 & 56.66 & 59.17
& 58.36 & 60.28 & 59.31
& 58.24 & 56.20 & 57.07
& 59.50 & 56.33 & 57.77
& 60.82 \\
& tiny
& 71.23 & 71.61 & 71.40
& 58.20 & 55.84 & 56.99
& 61.63 & 56.60 & 59.01
& 57.29 & 59.85 & 58.54
& 57.46 & 54.82 & 55.95
& 58.63 & 55.25 & 56.78
& 59.78 \\
\midrule
\multirow{3}{*}{Dasheng~\cite{Dasheng}}
& base
& 76.38 & 79.92 & 77.93
& 60.61 & 57.93 & 59.24
& 62.80 & 56.40 & 59.43
& 61.06 & 61.50 & 61.28
& 57.56 & 58.39 & 58.02
& 59.84 & 56.48 & 58.00
& 62.32 \\
& 0.6b
& 76.13 & 79.69 & 77.69
& 59.72 & 57.47 & 58.57
& 62.47 & 56.34 & 59.25
& 58.91 & 62.06 & 60.44
& 57.50 & 57.30 & 57.39
& 58.85 & 56.54 & 57.60
& 61.82 \\
& 1.2b
& 76.00 & 79.73 & 77.64
& 60.84 & 57.68 & 59.22
& 61.78 & 56.64 & 59.10
& 60.42 & 61.98 & 61.19
& 56.42 & 58.83 & 57.75
& 59.34 & 55.66 & 57.32
& 62.04 \\
\midrule
\multirow{2}{*}{EAT~\cite{EAT}}
& base
& 73.14 & 76.81 & 74.75
& 59.07 & 57.06 & 58.05
& 60.60 & 56.84 & 58.66
& 59.05 & 58.55 & 58.80
& 55.13 & 54.86 & 54.98
& 59.62 & 54.83 & 56.96
& 60.37 \\
& large
& 75.57 & 78.10 & 76.68
& 61.22 & 57.54 & 59.32
& 62.48 & 57.07 & 59.65
& 60.57 & 60.85 & 60.71
& 55.43 & 55.44 & 55.44
& 59.04 & 54.24 & 56.38
& 61.36 \\
\midrule
\multirow{3}{*}{FISHER~\cite{FISHER}}
& mini
& 72.68 & 77.87 & 74.96
& 60.44 & 57.33 & 58.84
& 59.90 & 55.45 & 57.59
& 57.63 & 61.15 & 59.34
& 53.86 & 55.36 & 54.69
& 59.01 & 53.86 & 56.15
& 60.26 \\
& small
& 72.69 & 79.02 & 75.47
& 61.29 & 58.97 & 60.11
& 64.34 & 56.48 & 60.15
& 59.06 & 62.38 & 60.68
& 56.06 & 55.49 & 55.74
& 58.57 & 55.58 & 56.93
& 61.51 \\
& tiny
& 71.72 & 77.91 & 74.44
& 60.23 & 56.67 & 58.39
& 60.52 & 55.97 & 58.16
& 58.88 & 56.86 & 57.85
& 54.28 & 57.78 & 56.20
& 59.22 & 53.92 & 56.27
& 60.22 \\
\midrule
\multirow{2}{*}{ECHO~\cite{ECHO}}
& small
& 77.06 & \underline{82.62} & 79.51
& \textbf{63.73} & \textbf{59.52} & \textbf{61.55}
& \underline{65.41} & 57.57 & \underline{61.24}
& \textbf{62.26} & \underline{67.40} & \underline{64.73}
& 57.87 & \underline{60.95} & \underline{59.56}
& \textbf{60.24} & \textbf{57.99} & \textbf{59.02}
& \underline{64.27} \\
& tiny
& 73.85 & 79.59 & 76.37
& 62.14 & 57.73 & 59.85
& 64.63 & 56.60 & 60.35
& \underline{61.63} & 67.30 & 64.34
& 55.94 & 57.91 & 57.03
& 58.92 & 57.64 & 58.23
& 62.69 \\
\midrule
\multirow{1}{*}{ECHOv2}
& small
& \textbf{77.64} & 82.16 & \underline{79.62}
& \underline{63.49} & \underline{59.39} & \underline{61.37}
& \textbf{65.62} & \textbf{58.65} & \textbf{61.94}
& 61.18 & \textbf{69.18} & \textbf{64.93}
& \underline{59.14} & \textbf{61.28} & \textbf{60.33}
& \underline{60.07} & \underline{57.95} & \underline{58.92}
& \textbf{64.52} \\
\bottomrule
\end{tabular}
}
\end{sidewaystable}

\begin{table}[t]
\centering
\caption{Ablation study of the inter-band learning design under the embedding-based ASD protocol. ``Overall'' denotes the arithmetic mean over DCASE 2020--2025. Best results in each column are marked in bold.}
\label{tab:ablation_interband_embed}
\begin{tabular}{l|ccccccc}
\toprule
\textbf{Method}
& \textbf{2020} & \textbf{2021} & \textbf{2022} & \textbf{2023} & \textbf{2024} & \textbf{2025} & \textbf{Overall} \\
\midrule
ECHOv2
& \textbf{72.96} & 60.36 & \textbf{60.48} & 63.55 & \textbf{58.80} & \textbf{59.62} & \textbf{62.63} \\
\hspace{0.8em}w/o inter-context
& 71.92 & 60.17 & 60.20 & 63.40 & 58.44 & 58.57 & 62.11 \\
\hspace{0.8em}w/o inter-reconstruction
& 72.90 & 60.34 & 60.01 & 63.20 & 58.29 & 58.90 & 62.27 \\
\hspace{0.8em}w/o FPE
& 72.89 & \textbf{60.46} & 60.15 & 63.51 & 58.53 & 59.26 & 62.46 \\
ECHO-Small~\cite{ECHO}
& 72.23 & 60.20 & 59.96 & \textbf{63.71} & 57.86 & 58.70 & 62.11 \\
\bottomrule
\end{tabular}
\end{table}

\begin{table}[t]
\centering
\caption{Ablation study of the inter-band learning design under the adaptation-based ASD protocol. ``Overall'' denotes the arithmetic mean over DCASE 2020--2025. Best results in each column are marked in bold.}
\label{tab:ablation_interband_adapt}
\begin{tabular}{l|ccccccc}
\toprule
\textbf{Method}
& \textbf{2020} & \textbf{2021} & \textbf{2022} & \textbf{2023} & \textbf{2024} & \textbf{2025} & \textbf{Overall} \\
\midrule
ECHOv2
& 79.62 & 61.37 & \textbf{61.94} & \textbf{64.93} & \textbf{60.33} & 58.92 & \textbf{64.52} \\
\hspace{0.8em}w/o inter-context
& 78.85 & 60.96 & 61.68 & 63.98 & 59.03 & 58.62 & 63.85 \\
\hspace{0.8em}w/o inter-reconstruction
& 79.44 & 61.09 & 61.29 & 64.36 & 59.86 & 58.95 & 64.17 \\
\hspace{0.8em}w/o FPE
& \textbf{79.82} & 61.45 & 61.45 & 64.72 & 60.02 & \textbf{59.29} & 64.46 \\
ECHO-Small~\cite{ECHO}
& 79.50 & \textbf{61.55} & 61.24 & 64.73 & 59.56 & 59.02 & 64.27 \\
\bottomrule
\end{tabular}
\end{table}

\begin{table}[t]
\centering
\caption{Effect of structured inter-band modeling with different numbers of summary tokens under the embedding-based ASD protocol. Here, the number of summary tokens $M$ controls the granularity of cross-band supervision in the inter-band branch. ``Overall'' denotes the arithmetic mean over DCASE 2020--2025. Best results in each column are marked in bold.}
\label{tab:ablation_sumtok_embed}
\begin{tabular}{c|ccccccc}
\toprule
\makecell{\textbf{\# Summary}\\\textbf{Tokens ($M$)}}
& \textbf{2020} & \textbf{2021} & \textbf{2022} & \textbf{2023} & \textbf{2024} & \textbf{2025} & \textbf{Overall} \\
\midrule
1
& \textbf{72.81} & 60.24 & 60.02 & 63.40 & 58.06 & 58.91 & 62.24 \\
3 (ECHOv2)
& 72.96 & \textbf{60.36} & \textbf{60.48} & 63.55 & \textbf{58.80} & \textbf{59.62} & \textbf{62.63} \\
6
& 72.65 & 60.32 & 60.15 & 63.14 & 58.11 & 58.57 & 62.16 \\
-- (ECHO-Small~\cite{ECHO})
& 72.23 & 60.20 & 59.96 & \textbf{63.71} & 57.86 & 58.70 & 62.11 \\
\bottomrule
\end{tabular}
\end{table}

\begin{table}[t]
\centering
\caption{Effect of structured inter-band modeling with different numbers of summary tokens under the adaptation-based ASD protocol. Here, the number of summary tokens $M$ controls the granularity of cross-band supervision in the inter-band branch. ``Overall'' denotes the arithmetic mean over DCASE 2020--2025. Best results in each column are marked in bold.}
\label{tab:ablation_sumtok_adapt}
\begin{tabular}{c|ccccccc}
\toprule
\makecell{\textbf{\# Summary}\\\textbf{Tokens ($M$)}}
& \textbf{2020} & \textbf{2021} & \textbf{2022} & \textbf{2023} & \textbf{2024} & \textbf{2025} & \textbf{Overall} \\
\midrule
1
& 79.12 & 61.15 & 61.35 & 64.32 & 59.47 & 58.77 & 64.03 \\
3 (ECHOv2)
& \textbf{79.62} & 61.37 & \textbf{61.94} & \textbf{64.93} & \textbf{60.33} & 58.92 & \textbf{64.52} \\
6
& 79.47 & 61.07 & 61.44 & 64.50 & 59.05 & 58.40 & 63.99 \\
-- (ECHO-Small~\cite{ECHO})
& 79.50 & \textbf{61.55} & 61.24 & 64.73 & 59.56 & \textbf{59.02} & 64.27 \\
\bottomrule
\end{tabular}
\end{table}

\subsection{Main Results on ASD Benchmarks}
\subsubsection{Results under the Embedding-based Protocol}
Table~\ref{tab:asd_main_raw} presents the ASD results under the embedding-based protocol on DCASE 2020--2025. Under this frozen-representation setting, ECHO-Small already provides a strong band-splitting baseline, achieving the best overall score among the compared baselines (62.11), ahead of BEATs (61.86) and EAT-Large (61.58). This result indicates that band-structured representations remain highly competitive for unified evaluation across multiple DCASE years.

On top of this strong baseline, ECHOv2 further improves the overall score from 62.11 to 62.63. 
The improvement is consistent across most benchmark years rather than being driven by a single favorable setting. 
More specifically, ECHOv2 outperforms ECHO-Small on DCASE 2020, 2021, 2022, 2024, and 2025, while remaining comparable on DCASE 2023.
This year-wise pattern indicates that the improvement is observed across heterogeneous ASD settings.

Such consistency is meaningful because DCASE 2020--2025 ASD tasks do not constitute identical evaluation setups. Across years, the benchmark varies in machine categories, operating conditions, domain-shift characteristics, and official aggregation rules. Therefore, the fact that ECHOv2 improves the overall average while yielding gains on most yearly benchmarks supports the view that the benefit of inter-band learning is not tied to a specific dataset configuration, but generalizes across heterogeneous ASD scenarios.

From the perspective of representation learning, these results suggest that inter-band supervision complements the ECHO intra-band training paradigm by injecting cross-frequency contextual cues into the learned band representations during pre-training.
Rather than altering the backbone architecture, ECHOv2 improves the representation through additional training-time supervision.
The consistent improvement over the ECHO baseline indicates that explicit inter-band modeling can provide additional supervisory value beyond the intra-band self-distillation scheme.

\subsubsection{Results under the Adaptation-based Protocol}
Table~\ref{tab:asd_main_adapt} reports the ASD results under the adaptation-based protocol. 
Compared with the embedding-based setting, this protocol evaluates frozen representations after a lightweight downstream adaptation module, providing a complementary view of representation transferability. 
Under this setting, ECHO-Small remains the strongest baseline, achieving the best overall score among the compared baselines (64.27). 
This result suggests that the band representation learned by ECHO is effective not only for direct embedding-based scoring, but also for downstream use with lightweight task adaptation.

ECHOv2 further improves the overall score from 64.27 to 64.52 and preserves the advantage over ECHO-Small under the adaptation-based protocol.
At the yearly level, ECHOv2 outperforms ECHO-Small on DCASE 2020, 2022, 2023, and 2024, while remaining very close on DCASE 2021 and 2025. 
This pattern indicates that the benefit of inter-band learning is largely preserved after downstream adaptation rather than being limited to the original frozen embedding space.

The adaptation-based results are also broadly consistent with those observed under the embedding-based protocol. 
Since the two protocols assess the learned representations from different angles---one through direct frozen-feature scoring and the other through a lightweight learned transformation---the consistent advantage of ECHOv2 under both settings suggests that inter-band learning improves representation quality in a more general sense, rather than only benefiting a specific evaluation pipeline.

Overall, the results under the two protocols support the same conclusion: ECHOv2 provides consistent gains over the ECHO baseline under both evaluation protocols. 
While the margins are moderate, they are consistently obtained over a strong band-splitting baseline, which supports the effectiveness and transferability of the proposed cross-band learning strategy.

\begin{table}[t]
\centering
\caption{Paired statistical significance analysis between ECHOv2 and representative strong baselines. The reported values are two-sided paired $t$-test $p$-values computed from year-level scores. ``Combined'' uses 12 paired observations from the two evaluation protocols.}
\label{tab:statistical_significance}
\begin{tabular}{lccc}
\hline
\textbf{Comparison} & \textbf{Embedding} & \textbf{Adaptation} & \textbf{Combined} \\
\hline
ECHOv2 vs. ECHO-Small & 0.0353 & 0.1843 & 0.0096 \\
ECHOv2 vs. FISHER-Small & 0.0101 & 0.0042 & $1.3{\times}10^{-4}$ \\
ECHOv2 vs. BEATs & 0.3033 & 0.0213 & 0.0139 \\
ECHOv2 vs. EAT-Large & 0.1469 & 0.0011 & 0.0012 \\
\hline
\end{tabular}
\end{table}

\subsection{Statistical Significance Analysis}
\label{subsec:statistical_significance}
Since the numerical margins among strong pre-trained audio backbones can be small, we further conduct two-sided paired $t$-tests using the year-level scores from DCASE 2020--2025.
We select representative methods from two categories: structurally related band-splitting baselines, including ECHO-Small and FISHER-Small, and strong general-purpose pre-trained audio backbones according to the overall scores in the main result tables, including BEATs and EAT-Large.
We compare ECHOv2 with these representative baselines under the embedding-based and adaptation-based protocols.
We also report a combined test over the 12 year-protocol pairs as an additional summary.

Table~\ref{tab:statistical_significance} shows that ECHOv2 significantly improves over ECHO-Small under the embedding-based protocol ($p=0.0353$) and in the combined test ($p=0.0096$), while the adaptation-based gain over ECHO-Small is positive but not significant at the 0.05 level.
Compared with FISHER-Small, ECHOv2 shows significant improvements under both protocols.
For BEATs, the improvement is not significant under the embedding-based protocol but becomes significant under the adaptation-based protocol and in the combined test, suggesting that the advantage of ECHOv2 is more evident after lightweight downstream adaptation.
For EAT-Large, ECHOv2 shows significant improvement under the adaptation-based protocol and in the combined test.
These results provide paired statistical evidence that the overall gains of ECHOv2 are generally consistent across benchmark years and evaluation protocols.

\subsection{Ablation Study}
\subsubsection{Effect of Inter-Band Learning}
The ablation results in Tables~\ref{tab:ablation_interband_embed} and~\ref{tab:ablation_interband_adapt} analyze the contribution of each component in the inter-band branch under the embedding-based and adaptation-based protocols, respectively.
Among the examined components, the inter-context objective provides the largest contribution, while inter-reconstruction and FPE offer additional refinements.

A particularly clear pattern appears when the inter-context objective is removed. 
Under the embedding-based protocol, the overall score drops from 62.63 to 62.11, exactly returning to the level of ECHO-Small. 
Under the adaptation-based protocol, the score further decreases from 64.52 to 63.85, which is even lower than the ECHO-Small baseline (64.27). 
This behavior indicates that the effectiveness of the inter-band branch mainly comes from the contextual supervision imposed on cross-band aggregation.

The effect of inter-reconstruction is more moderate. 
Without this objective, the model still reaches 62.27 under the embedding-based protocol, which remains above ECHO-Small, indicating that inter-context supervision alone already accounts for a substantial part of the improvement. 
Under the adaptation-based protocol, however, the score decreases to 64.17, slightly below the ECHO-Small result. 
This suggests that inter-reconstruction is useful, but its contribution is better understood as supportive rather than dominant. 
One possible explanation is that the adaptation-based protocol may be more sensitive to the global organization of the representation space, where explicit inter-context supervision becomes particularly useful.

Removing FPE causes only small changes in the overall results, yielding 62.46 and 64.46 under the two protocols. 
This suggests that FPE acts as a lightweight refinement that provides positional guidance during cross-band aggregation, rather than serving as the main source of improvement.

Taken together, the ablation shows that the inter-band branch is driven primarily by inter-context supervision, with inter-reconstruction offering additional support and FPE acting as a lightweight refinement.

\subsubsection{Influence of Summary Token Granularity}
Tables~\ref{tab:ablation_sumtok_embed} and~\ref{tab:ablation_sumtok_adapt} study how the number of summary tokens affects the structured inter-band modeling design under the embedding-based and adaptation-based protocols, respectively.
In our formulation, the token number $M$ determines the granularity of cross-band aggregation in the inter-band branch: $M=1$ corresponds to a single global aggregation pathway, while larger $M$ allows the model to organize cross-band interactions in a more structured manner.

A consistent trend can be observed under both evaluation protocols. 
Moving from one summary token to three improves the overall score from 62.24 to 62.63 in the embedding-based setting and from 64.03 to 64.52 in the adaptation-based setting. 
This indicates that a single global summary may be too coarse to capture the diversity of cross-band dependencies involved in machine sound analysis. 
In contrast, increasing the number further to six does not bring additional gains, with the overall results dropping to 62.16 and 63.99, respectively.

The year-wise results follow the same general pattern. 
The three-token configuration achieves the best overall performance and attains the strongest result in most yearly evaluations across the two protocols, whereas the one-token and six-token settings are generally less competitive. 
Rather than indicating that more summary tokens are always preferable, these results point to the importance of choosing an appropriate aggregation granularity for inter-band modeling.

From this perspective, the three-token setting appears to offer a better balance between global integration and structured frequency-region modeling. 
Using only one summary token tends to enforce overly coarse aggregation, while using too many tokens does not translate into better downstream ASD performance. 
The results therefore support the use of a moderate number of summary tokens as a practical design choice for structured inter-band learning.

\begin{table}[t]
\centering
\caption{Representation analysis with different numbers of summary tokens. 
``sumtok'' denotes the concatenated summary-token representation, and ``+'' denotes feature concatenation. 
``Average'' is the arithmetic mean of the embedding- and adaptation-based ASD scores.}
\label{tab:representation_analysis}
\begin{tabular}{cc|ccc}
\toprule
M & \textbf{Representation}
& \makecell{\textbf{Embedding-based}\\\textbf{ASD}}
& \makecell{\textbf{Adaptation-based}\\\textbf{ASD}}
& \textbf{Average}\\
\midrule
\multirow{3}{*}{1}
& band          & 62.24 & 64.03 & \textbf{63.14} \\
& sumtok        & 59.47 & 59.29 & 59.38 \\
& band+sumtok   & 62.33 & 62.92 & 62.63 \\
\midrule
\multirow{3}{*}{3}
& band          & 62.63 & 64.52 & \textbf{63.58} \\
& sumtok        & 60.60 & 60.61 & 60.61 \\
& band+sumtok   & 62.68 & 63.11 & 62.90 \\
\midrule
\multirow{3}{*}{6}
& band          & 62.16 & 63.99 & \textbf{63.08} \\
& sumtok        & 62.50 & 62.66 & 62.58 \\
& band+sumtok   & 62.21 & 62.87 & 62.54 \\
\bottomrule
\end{tabular}
\end{table}

\subsection{Discussion on Downstream Representations}

\noindent
ECHOv2 is designed to improve the ECHO band representation through explicit cross-band supervision while preserving the band-splitting representation format for downstream ASD.
At the same time, the introduction of structured summary tokens raises a natural question: beyond supporting inter-band learning during training, can these additional representations also serve as effective downstream features for ASD? 
To address this question, Table~\ref{tab:representation_analysis} compares the band representation, the summary-token representation, and their concatenation.

\subsubsection{Band vs. Summary-Token Representations}
The band representation remains the most reliable downstream feature across the evaluated settings. 
Under the default configuration of $M=3$, it achieves the highest average score, reaching 63.58 across the embedding-based and adaptation-based protocols. 
This supports our default choice of using the final band-level representation for downstream ASD evaluation.

A likely reason is that the band representation preserves band-specific structure while also benefiting from the cross-band contextual refinement introduced by inter-band learning. 
By contrast, the summary-token representation is more compact and more strongly oriented toward aggregated global information, which makes it less effective at retaining the fine-grained spectral cues needed for ASD.

\subsubsection{Summary Tokens as Structured Aggregation Units}
Used alone, summary-token representations are generally weaker than band representations, especially when $M=1$ or $M=3$. 
This indicates that summary tokens are useful for aggregating cross-band information, but are not the most suitable standalone representation for downstream inference.

Their behavior nevertheless remains informative. 
As the number of summary tokens increases, the average score of the summary-token representation rises from 59.38 for $M=1$ to 60.61 for $M=3$, and further to 62.58 for $M=6$. 
This trend suggests that finer structured aggregation increases the expressive capacity of the summary-token pathway itself. 
Even so, the band representation remains stronger in all three settings. 
The role of summary tokens is therefore better understood as facilitating structured cross-band interaction during training rather than replacing the final band representation at inference time.

\subsubsection{Is Concatenation Necessary?}
The concatenation results lead to a similar conclusion. 
Although band+sumtok consistently outperforms the summary-token representation alone, it does not provide a clear benefit over the band representation by itself. 
This is most evident under the adaptation-based protocol, where the concatenated representation is lower than the band representation for all three values of $M$.

This pattern suggests that much of the useful information captured by the summary-token pathway has already been incorporated into the learned band representation through inter-band learning. 
Once the band-level feature has been refined in this way, explicit concatenation adds little for downstream ASD.

Overall, the evidence points to a clear division of roles in ECHOv2: the band representation remains the most stable and effective downstream feature, whereas summary tokens mainly serve as structured aggregation units that support inter-band learning during training.

\section{Limitations and Future Work}
\label{sec:limitations}

Although ECHOv2 achieves strong performance on the unified DCASE 2020--2025 ASD benchmark, several directions remain for future investigation.
First, the current benchmark experiments are conducted on the official DCASE Task~2 datasets from 2020 to 2025.
While these datasets provide a reproducible basis for multi-year ASD evaluation, applying the benchmark to additional ASD datasets still requires dataset-specific adaptation of metadata parsing, reference-bank construction, and scoring definitions.
Second, ECHOv2 uses a fixed sub-band partition and a fixed number of summary tokens during pre-training.
Different machine types or operating conditions may benefit from adaptive frequency grouping or data-dependent summary-token allocation.
Third, this work focuses on representation-level evaluation with frozen backbones and lightweight downstream adaptation.
Future work may further integrate ECHOv2 with more task-specific ASD scoring and adaptation strategies while preserving the benefits of frequency-structured representation learning.

\section{Conclusion}
This paper presented ECHOv2, a frequency-structured pre-trained acoustic representation model for machine anomalous sound detection.
ECHOv2 learns localized intra-band representations while introducing explicit inter-band supervision through a two-level self-distillation strategy.
By combining inter-band context alignment, masked sub-band reconstruction, and structured multi-summary-token aggregation, ECHOv2 captures both local spectral structure and cross-frequency dependencies.
We also established a unified ASD benchmark over DCASE 2020--2025 with two complementary protocols: embedding-based evaluation for frozen representation discriminability and adaptation-based evaluation for downstream transferability.
Experimental results, ablation studies, and statistical significance analysis demonstrate the effectiveness of ECHOv2 as a transferable acoustic representation model for industrial ASD.
The model and benchmark are fully open-sourced to support reproducible research.

\section*{Acknowledgements}

This research is funded in part by the National Natural Science Foundation of China (62571223) and the Science and Technology Program of Suzhou City (SYC2022051). Many thanks for the computational resource provided by the Advanced Computing East China Sub-Center.

\section*{Generative AI statement}

During the preparation of this manuscript, the authors used generative AI tools for limited language editing purposes, including improving clarity and correcting grammatical issues. No substantive content, analysis, or conclusions were generated by AI. The authors remain fully responsible for the content of this manuscript.

\section*{Declaration of Competing Interest}

The authors declare that they have no known competing financial interests or personal relationships that could have appeared to influence the work reported in this paper.

\printcredits

\bibliographystyle{cas-model2-names}

\bibliography{cas-refs}

\end{document}